\documentclass[twocolumn,showpacs,preprintnumbers,amsmath,amssymb
 aps,
 prb,
 lengthcheck,%
]{revtex4-1}
\usepackage{amssymb}
\usepackage{graphicx}
\usepackage{dcolumn}
\usepackage{bm}
\usepackage{hyperref}
\usepackage{makeidx}

\usepackage[english]{babel} 
\usepackage[latin1]{inputenc}
\usepackage{color}
\usepackage{graphicx}
\usepackage{amsmath}

\usepackage{latexsym}
\usepackage{amsthm}

\makeindex
\def\>{\right\rangle}
\def\<{\left\langle}
\def\be{\begin{equation}}
\def\ee{\end{equation}}
\def\ba{\begin{array}{l}}
\def\ea{\end{array}}

\def\beq{\begin{eqnarray}}
\def\eeq{\end{eqnarray}} 

\begin{document}

\preprint{APS/123-QED}
 
\title{Enhancing photon squeezing one Leviton at a time.}

\author{D. Ferraro$^{1}$, F. Ronetti$^{2, 3, 4}$, J. Rech$^{4}$, T. Jonckheere$^{4}$, M. Sassetti$^{2,3}$ and T. Martin$^{4}$ }
\affiliation{$^1$ Istituto Italiano di Tecnologia, Graphene Labs, Via Morego 30, I-16163 Genova, Italy\\
$^2$ Dipartimento di Fisica, Universit\`a di Genova, Via Dodecaneso 33, 16146, Genova, Italy\\
$^3$ CNR-SPIN, Via Dodecaneso 33, 16146, Genova, Italy\\
$^4$ Aix Marseille Univ, Universit\'e de Toulon, CNRS, CPT, Marseille, France\\}

\date{\today}

\begin{abstract}
A mesoscopic device in the simple tunnel junction or quantum point contact geometry emits microwaves with remarkable quantum properties, when subjected to a sinusoidal drive in the GHz range. In particular, single and two-photon squeezing as well as entanglement in the frequency domain have been reported. By revising the photo-assisted noise analysis developed in the framework of electron quantum optics, we present a detailed comparison between the cosine drive case and other experimentally relavent periodic voltages such as rectangular and Lorentzian pulses. We show that the latter drive is the best candidate in order to enhance quantum features and purity of the outgoing single and two-photon states, a noteworthy result in a quantum information perspective. 
\end{abstract}

\pacs{72.70.+m, 73.23.-b, 42.50.-p}
\maketitle
\section{Introduction}
The rise of electron quantum optics (EQO) \cite{grenier_electron_2011,bocquillon_electron_2014} as a prominent topic in the condensed matter agenda has been possible thanks to a remarkable synergy between experimental observations and theoretical investigations. In particular, the realization of on-demand electron sources based on periodically driven quantum dots \cite{feve_-demand_2007} or trains of Levitons, \emph{i.~e.} properly designed Lorentzian voltage pulses in time \cite{dubois_minimal_2013, jullien_quantum_2014}, represented the actual starting point of this new field of research. They allow the on-demand injection of individual excitations into mesoscopic devices mimicking the conventional photonic quantum optics with quantum Hall edge channels behaving as waveguides and quantum point contacts (QPC) playing the role of half-silvered mirrors. 

A milestone of this branch of mesoscopic physics has been the electronic translation of few-excitation interferometers like: the Hanbury-Brown-Twiss \cite{hanburybrown_test_1956}, able to access the granular nature of the particles through partitioning at a QPC, and the Hong-Ou-Mandel \cite{hong_measurement_1987}, where the statistical properties of the excitations are investigated by means of controlled two-particle collisions. Measurements clearly showed anti-bunching effects related to the fermionic nature of the electrons \cite{bocquillon_electron_2012a} as well as dephasing and decoherence induced by the electron-electron interaction \cite{bocquillon_coherence_2013, wahl_interactions_2014, ferraro_real_2014}, remarkable phenomena without parallel in the photonic domain.
 
 New experimental investigations carried out in tunnel junctions \cite{gasse_observation_2013, forgues_experimental_2015, Forgues16} and Josephson junctions \cite{Westig17} have shown the deep connection between the finite frequency photo-assisted noise generated by a periodic drive applied to a mesoscopic device and the fluctuations of the corresponding emitted radiation in the microwave regime. \cite{beenakker_counting_2001, beenakker_antibunched_2004, grimsmo_quantum_2016, Mora15b} In particular, measurements show unequivocally that the outgoing radiation is strongly non-classical presenting quantum features such as: single-photon squeezing \cite{gasse_observation_2013}, two-photon squeezing and entanglement in the frequency domain. \cite{forgues_experimental_2015} These results naturally opened interesting perspectives in the quantum information domain. \cite{braunstein_quantum_2016} However, this phenomenology has been investigated so far only for the cases of harmonic and bi-harmonic drives \cite{Gabelli17} and a more detailed analysis of possible voltage profiles in view of optimizing the quantum properties of the emitted radiation is still lacking. 

In this paper, taking advantage of the results achieved \cite{dubois_minimal_2013} and the tools \cite{dubois_integer_2013, rech_minimal_2016,Vannucci17,Ronetti17} elaborated in the framework of EQO, we compare  the photo-assisted finite frequency noise associated with the current outgoing from a non-interacting tunnel junction after applying an experimentally realizable drive like cosine, rectangular and Lorentzian signals, showing that the latter voltage represents the optimal compromise between the maximization of the squeezing and of the purity of the emitted single and two-photon states. 

This provides the proper theoretical framework to realize new experiments devoted to controlling and improving the quantum behavior of the associated emitted electromagnetic radiation. Our analysis also contributes to start a new phase of EQO. Indeed, until now, the main motivation behind this branch of physics has been to properly revise conventional optics experiments for excitations propagating ballistically in condensed matter systems. With this work, we shed light on the consequences of optimal injection of individual electron wave-packets on the quantum properties of the emitted photons.

The paper is organized as follows. In Section \ref{Quadratures} we discuss the connections between the current fluctuations and the quadratures of the emitted electromagnetic field. A possible measurement set-up able to access these fluctuations is shown in Section \ref{Setup}. Section \ref{Calculation} is devoted to the quantum mechanical calculation of the photo-assisted noise at finite frequency associated with various possible voltage drives. The characterization of the single-photon squeezing as well as the purity of the states generated by the different drives are discussed in Section \ref{single_photon}. In Section \ref{Two_photons} we investigate the features associated with two-photon states and characterize their entanglement. Finally, Section \ref{Conclusions} is devoted to the conclusions. 

\section{Quadratures of the emitted electromagnetic field}\label{Quadratures} According to Ref. \onlinecite{grimsmo_quantum_2016}, the current operator $I$ describing the charge flowing through a mesoscopic device can be connected to the outgoing electromagnetic field annihilation operator $a$ through the relation 
\be
a(\omega)= -i \frac{I(\omega)}{\sqrt{2 \mathcal{A}(\omega)}}
\ee
with $\mathcal{A}(\omega)=G F \hbar \omega$ a function which depends linearly on: the linear conductance  $G$ of the tunnel junction (assumed here energy independent), the Fano factor $F$ and the fixed measurement frequency $\omega$ of the detection set-up (see below). 

Notice that the above equation exactly holds only in the case of ideal matching at low impedance between the sample and the measurement set-up \cite{grimsmo_quantum_2016}. 

The quadratures of the electromagnetic field are then defined as \cite{gasse_observation_2013}
\beq
A(\omega)&=&\frac{1}{\sqrt{2}}\left[I(\omega) +I(-\omega)\right]\nonumber \\
&=&i \sqrt{\mathcal{A}(\omega)} \left[a(\omega)-a^{\dagger}(\omega) \right]\\
B(\omega)&=&\frac{i}{\sqrt{2}}\left[I(\omega) -I(-\omega)\right]\nonumber\\
&=&-\sqrt{\mathcal{A}} \left[a(\omega)+a^{\dagger}(\omega) \right]
\eeq
Using the Robertson formulation of the Heisenberg principle 
\be
\Delta A \Delta B\geq \frac{1}{2} |\langle\left[A, B\right] \rangle |,
\ee
with $\Delta A=\sqrt{\langle A^{2}\rangle-\langle A \rangle^{2}}$ (and an analogous expression for the operator $B$) and $\left[A,B\right]$ the usual definition of the commutator, as well as the conventional bosonic commutation relation
\be
\left[ a(\omega), a^{\dagger}(\omega)\right]=1,
\ee
one directly obtains 
\be
\Delta A \Delta B\geq \mathcal{A}
\ee
which naturally connects the quantum fluctuations of the electromagnetic field quadrature at a given frequency $\omega$ with the current fluctuations, namely the finite frequency noise. An experimental scheme to detect these fluctuations will be presented in the following Section. 

\section{Experimental set-up}\label{Setup}
\begin{figure}[h]
\centering
\includegraphics[scale=0.35]{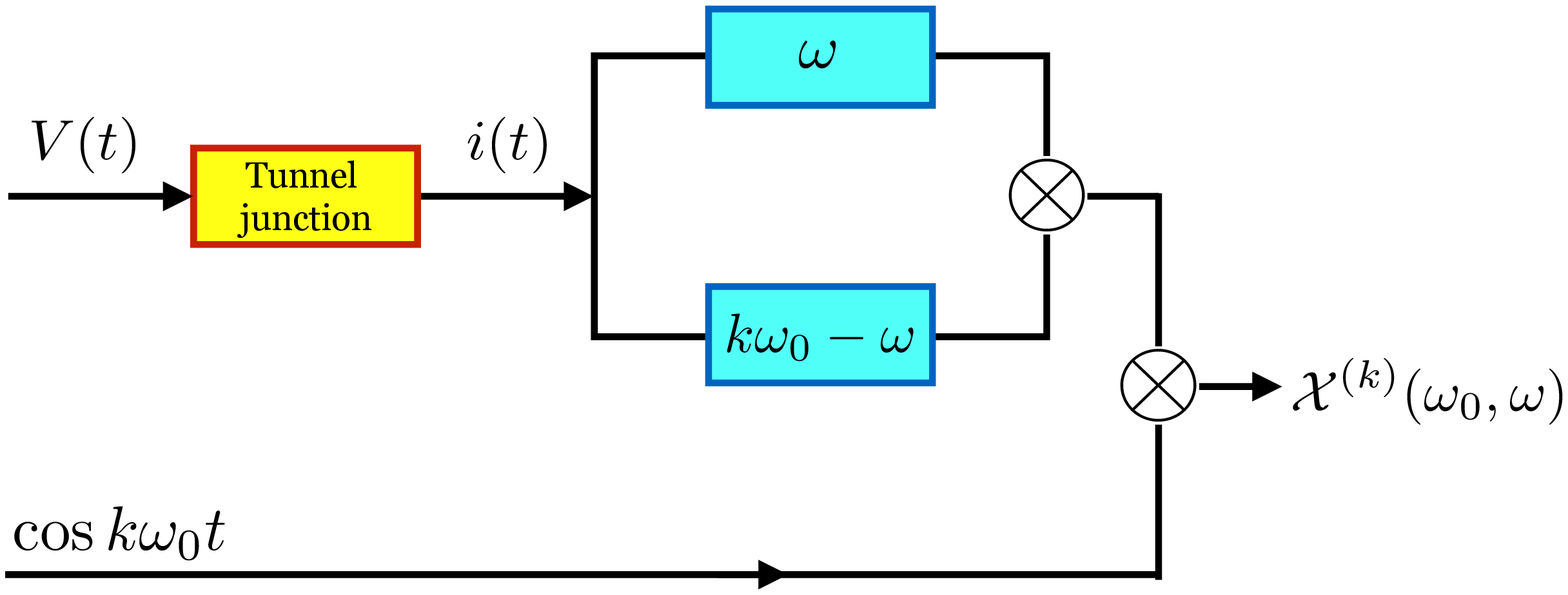}
\caption{(Color on-line) Schematic view of a two-filters set-up designed to measure the dynamical response of the noise $\mathcal{X}^{(k)}$. A tunnel junction (yellow box) is driven by a time dependent voltage $V(t)$ and emits a current $i(t)$ which is split into two. The resulting contributions are filtered at frequency $\omega$ and $|k \omega_{0}-\omega|$ respectively (blue boxes). They are then multiplied among themselves and with a cosine signal ($\otimes$ symbols). The final output is then averaged over a period.}
\label{fig1}
\end{figure}  

When a mesoscopic device is subjected to an external non-adiabatic AC drive at frequency $\omega_{0}$ (in the GHz range) a two-filter measurement (see Fig. \ref{fig1}) allows to access both the stationary photo-assisted noise at finite frequency \cite{blanter_shot_2000, martin_noise_2005, lesovik_noise_1994, schoelkopf_observation_1998} (zero-th order harmonic) and a more general dynamical response of the current fluctuations \cite{gabelli_dynamics_2008, gabelli_noise_2008}, corresponding to the higher order harmonics at a frequency $k\omega_{0}$ ($k \in \mathbb N^{*}$).

After filtering at a given frequency $\Omega$, the physical current $i(t)$ outgoing from the sample becomes 
\be
i(t)\rightarrow i_{\Omega}(t)\approx  i (\Omega) e^{i \Omega t}+ i (-\Omega) e^{-i \Omega t},   
\ee
providing an \emph{operative} definition for the $k$-th harmonics of the dynamical response of the noise measured at a frequency $\omega$ (again in the GHz range)
\be
\mathcal{X}^{(k)}(\omega_{0}, \omega)= \langle \overline{i_{\omega}(t) i_{k\omega_{0}-\omega}(t) \cos(k \omega_{0}t)} \rangle,
\label{X_k_generic} 
\ee
according to the functioning of the set-up described in Fig. \ref{fig1}. \cite{gabelli_noise_2008}  Notice that the filtering procedure consists in extracting the DC part by time averaging (operation denoted by $\overline{\langle ... \rangle}$). Therefore, keeping only the non zero terms in the average, the expression in Eq. (\ref{X_k_generic}) reduces to
\be
\mathcal{X}^{(k)}(\omega_{0}, \omega)= \frac{1}{2} \left[\langle i (\omega) i(k \omega_{0}-\omega)\rangle+\langle i (-\omega) i(-k\omega_{0}+\omega) \rangle\right].
\label{exp_definition}
\ee
 
\section{Quantum mechanical calculation}\label{Calculation}
In order to theoretically investigate the above quantity, we need to link the definition in Eq. (\ref{exp_definition}) with a \emph{quantum mechanical} one by introducing symmetrized correlators of current operators, namely \cite{gabelli_noise_2008, ferraro_multiple_2014}
\be
\langle i (\omega_{1}) i (\omega_{2})\rangle\Rightarrow \frac{1}{2} \left[\langle I(\omega_{1}) I(\omega_{2})\rangle_{c}+\langle I(\omega_{2}) I(\omega_{1})\rangle_{c} \right]. 
\ee 
We want to remark here the fact that, for sake of clarity, we distinguished the notation for the physical current ($i$) and the associated quantum mechanical operator ($I$). Moreover, we introduced the notation $\langle \mathcal{O}_{1} \mathcal{O}_{2}\rangle_{c}=\langle \mathcal{O}_{1} \mathcal{O}_{2}\rangle-\langle \mathcal{O}_{1} \rangle\langle \mathcal{O}_{2}\rangle$ for quantum mechanical correlator (at zero or finite temperature $T$) between two arbitrary operators $\mathcal{O}_{1}$ and $\mathcal{O}_{2}$. According to this, and taking into account the reality condition $I^{\dagger}(\omega)=I(-\omega)$ for the current operator, one can write  
\be
\mathcal{X}^{(k)}(\omega_{0}, \omega)= \frac{1}{2} \left[\Re\left\{\mathcal{X}^{(k)}_{+}(\omega_{0}, \omega)\right\}+\Re\left\{\mathcal{X}^{(-k)}_{+}(\omega_{0}, -\omega)\right\}\right] \label{Xp}
\ee
with
\be
\mathcal{X}^{(k)}_{+}(\omega_{0}, \omega)=\langle I(\omega) I(k\omega_{0}-\omega) \rangle_{c}
\ee
and $\Re\left\{...\right\}$ indicating the real part.
 
We can consider now the experimentally relevant case of a non-interacting tunnel junction (or QPC geometry) subjected to a periodic drive $V(t)$ with period $\mathcal{T}=2 \pi /\omega_{0}$, which can be naturally decomposed into DC and AC part as 
\be
V(t)=V_{DC}+V_{AC}(t)
\ee
with 
\begin{equation}
\frac{1}{\mathcal{T}}\int^{+\frac{\mathcal{T}}{2}}_{-\frac{\mathcal{T}}{2}} dt V_{AC}(t)=0.
\end{equation}

In this case one obtains the explicit form for the dynamical response of the noise at frequency $k \omega_{0}$ 
\beq
\mathcal{X}^{(k)}&=&\frac{1}{2} \Re\left\{\sum_{n=-\infty}^{+\infty} p_{n}(z) p^{*}_{n+k}(z) \mathcal{S}_{0}(q+1+n\lambda, \theta) \right.\nonumber\\
&+&\left.  \sum_{n=-\infty}^{+\infty} p_{n}(z) p^{*}_{n-k}(z) \mathcal{S}_{0}(q-1+n\lambda, \theta)  \right\}.\nonumber\\
\label{Xp_extended}
\eeq
Notice that, in the above expression we introduced the short notations: $q=e V_{DC}/\hbar \omega$, $\lambda=\omega_{0}/\omega$, $\theta=k_{B}T/\hbar \omega$ and $z=e \tilde{V}/\hbar \omega_{0}$ ($\tilde{V}$ the amplitude of the AC voltage) and we have omitted the functional dependence for notational convenience. 

The function  
\be
\mathcal{S}_{0}(\xi, \theta)= \mathcal{A} \xi \coth \left(\frac{\xi}{2 \theta} \right),
\label{Szero}
\ee 
represents the rescaled variables version of the thermal/shot noise crossover formula 
\be
\mathcal{S}_{0}(V_{DC}, T)= G F e V_{DC} \coth \left(\frac{e V_{DC}}{2 k_{B} T} \right)
\ee
evaluated for a tunnel junction geometry in a normal metal.\cite{blanter_shot_2000, martin_noise_2005}

 In the zero temperature limit it reduces to  
\be
\mathcal{S}_{0}(\xi, \theta \rightarrow 0)= \mathcal{A} |\xi|.
\ee
Moreover, according to the general definition of the photo-assisted amplitudes \cite{crepieux_photoassisted_2004, dubois_integer_2013}, one has
\begin{equation}
p_{n}(z) = \int_{-\mathcal{T}/2}^{\mathcal{T}/2} \frac{dt}{\mathcal{T}} e^{2 i \pi  n \frac{t}{\mathcal{T}}} e^{- 2 i\pi  z  \varphi(t)}, 
\label{eq:defpofl}
\end{equation}
with 
\be
\varphi (t) = e \int_{-\infty}^t \frac{dt'}{\mathcal{T}}\bar{V}_{AC}(t'),
\ee
where $\bar{V}_{AC}(t)$ is the AC part of $V(t)$ with unitary and dimensionless amplitude.\\
It is worth to note that for $k=0$ the expression in Eq. (\ref{Xp}) reduces to
\begin{widetext}
\be
\mathcal{X}^{(0)}=\tilde{\mathcal{S}}=\frac{1}{2} \sum_{n=-\infty}^{+\infty}P_{n}(z) \left[\mathcal{S}_{0}(q+1+n\lambda, \theta)+ \mathcal{S}_{0}(q-1+n\lambda, \theta)\right],
\label{S_tilde}
\ee
\end{widetext}
with $P_{n}(z)=|p_{n}(z)|^{2}$, and represents the photo-assisted noise $\tilde{\mathcal{S}}$ measured at finite frequency $\omega$, as expected.

In order to proceed with our analysis, we need to specify the functional form of the photo-assisted amplitudes $p_{n}(z)$ for experimentally relevant voltage drives. In the following we will focus on three different signals: a cosine, a rectangular periodic drive and a train of Lorentzian pulses. \cite{dubois_integer_2013} We will keep separate the DC and AC amplitude in such a way to consider the parameters $q$ (corresponding to the number of electrons injected in the junction by the drive \cite{dubois_integer_2013, rech_minimal_2016}) and $z$ in Eq. (\ref{Xp_extended}) as totally independent. 

\subsection{Cosine drive}
The simple sinusoidal drive we are considering is
\be
V_{AC}(t) = -\tilde{V} \cos \left( \omega_{0} t \right)
\ee
leading to photo-assisted amplitudes in Eq. (\ref{eq:defpofl}) of the form
\be
p_{n}(z) = J_{n} \left(-z\right).
\ee
Notice that here we introduced $n$-th Bessel's function of the first kind $J_{n}(x)$. They satisfy the relation 
\be
J_{-n}(x)= (-1)^{n} J_{n}(x)
\ee
which allows to further manipulate Eq. (\ref{Xp_extended}) in such a way to recover the expression reported in Refs. \onlinecite{gabelli_noise_2008, gasse_observation_2013}. 

\subsection{Rectangular drive}
The AC voltage profile in this case is given by 
\be
  V_{AC}(t) = \left\{
    \begin{array}{cc}
      -\tilde{V} & \text{for} -\frac{\mathcal{T}}{2} \leq t <-\eta\frac{\mathcal{T}}{2},\\
     \tilde{V}\left(\frac{1}{\eta}-1 \right) & \text{for} -\eta\frac{\mathcal{T}}{2} \leq t \leq \eta\frac{\mathcal{T}}{2} ,\\
      -\tilde{V} & \text{for }  \eta\frac{\mathcal{T}}{2} \leq t < \frac{\mathcal{T}}{2}
    \end{array} \right.
\ee
with $\eta$ the width of the pulse in units of the period. The associated photo-assisted probability amplitudes read
\be
p_{n}(z)= \frac{z \sin\left\{\pi \left[(\eta-1)z+\eta n \right]\right\}}{\pi (z+n)\left[\left(\eta-1 \right)z+\eta n\right]}.
\ee
Notice that this particular drive reduces to the conventional square wave for $\eta=1/2$ and that for $\eta \rightarrow 0$ (and $z=1/2$) one has
\be
p_{n}(z=\frac{1}{2})= \frac{2}{\pi \left(2n+1 \right)},
\label{Dirac_comb}
\ee
in agreement with what reported in Ref. \onlinecite{mendes_cavity_2015} for the case of a Dirac comb in time. Moreover, for $\eta=1$ the photo-assisted amplitudes are zero, as expected for a purely DC bias. 

\subsection{Lorentzian voltage pulses}
A periodic train of Lorentzian pulses has an AC contribution  

\be
V_{AC}(t) = \frac{\tilde{V}}{\pi} \sum_{l=-\infty}^{+\infty} \frac{\eta}{\eta^2 + \left( \frac{t}{\mathcal{T}}-l \right)^2}-\tilde{V}
\ee
characterized by the photo-assisted amplitudes \cite{dubois_integer_2013, grenier_fractionalization_2013}

\be
p_{n}(z) =  z \sum_{s=0}^{+\infty} \frac{\Gamma (z+n+s)}{\Gamma(z+1-s)} \frac{(-1)^s e^{-2 \pi \eta (2s+n)}}{(n+s)! s!},
\ee
with $\Gamma(x)$ the Euler's Gamma function. Notice that, in this case, $\eta$ parametrizes the width at half height of the pulse and in the limit $\eta\rightarrow 0$ we recover the Dirac comb case (see Eq. (\ref{Dirac_comb})).  

\section{Single-photon squeezing}\label{single_photon}
Due to the previous considerations about the current correlators one can write the quadratures fluctuations of the emitted radiation as \cite{gasse_observation_2013}
\beq
\langle A^{2} \rangle_{c}&=&(\Delta A)^{2}=\tilde{\mathcal{S}}+\mathcal{X}^{(k)}
\label{squeezing_noise1}\\
\langle B^{2} \rangle_{c}&=&(\Delta B)^{2}=\tilde{\mathcal{S}}-\mathcal{X}^{(k)}
\label{squeezing_noise2}
\eeq
where, as stated above, the squeezing of the emitted electromagnetic field is achieved for $\langle A^{2} \rangle_{c}/\mathcal{A}<1$ or $\langle B^{2} \rangle_{c}/\mathcal{A}<1$.
\begin{figure}[h]
\centering
\includegraphics[scale=0.28]{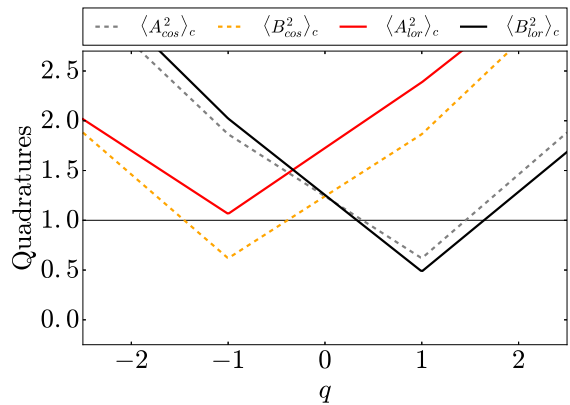}
\caption{(Color on-line) Quadratures of the emitted electromagnetic field in units of $\mathcal{A}$, as a function of $q$ (number of injected electrons) and at zero temperature ($\theta=0$). Curves represent: $\langle B^{2}_{lor} \rangle_{c}$ for the Lorentzian drive at $z=0.856$ and $\eta=0.1$ (full black curve), $\langle A^{2}_{lor} \rangle_{c}$ for the Lorentzian drive at $z=0.856$ and $\eta=0.1$ (red full curve), $\langle A^{2}_{cos} \rangle_{c}$ at $z=0.706$ for the cosine drive case (gray dashed curve) and $\langle B^{2}_{cos} \rangle_{c}$ at $z=0.706$ for the cosine drive case (orange dashed curve). Thin black horizontal line indicates the vacuum fluctuations (in units of $\mathcal{A}$). The choice of the parameter $z$ in each curve has been achieved by numerical minimization of the noise.} 
\label{fig2}
\end{figure}
We consider now the behavior of the quadratures for different drives measured at $\omega=\omega_{0}/2$ ($k=1$) and zero temperature ($\theta=0$), which has been already shown to be the more favorable configuration in order to enhance squeezing in the sinusoidal drive case. \cite{grimsmo_quantum_2016} It is shown in Fig. \ref{fig2} for drives which are in phase with the periodic current probe $\cos{k\omega_{0}t}$. Out of phases signals lead, in general, to a suppression of $\mathcal{X}^{(k)}$ and consequently of the squeezing effect (see Eq. (\ref{Xp_extended})). The presented results have been obtained by numerically minimizing the noise as a function of $z$ (amplitude of the AC drive). It is easy to note that for the cosine drive the fluctuations of the quadrature $\langle A^{2}_{cos} \rangle_{c}$ ($\langle B^{2}_{cos} \rangle_{c}$) goes below the quantum vacuum (thin black horizontal line) for $q=1$ ($q=-1$) at expenses of the other which stays well above, as already reported in Ref. \onlinecite{gasse_observation_2013}. This phenomenology is exactly what is expected for a generic squeezed state where the fluctuations along one quadrature are suppressed, while the ones along the canonically conjugated one are enhanced in order to preserve the Heisenberg principle. Very similar behavior is obtained for the square drive (rectangular pulse at $\eta=1/2$, not shown) even if in this case the noise signal is slightly above the one for the cosine drive at any temperature (see Table \ref{tab1}). The explicit mirror symmetry  connecting $\langle A^{2}_{cos} \rangle_{c}$ and $\langle B^{2}_{cos} \rangle_{c}$ (also present in the square drive) is in full agreement with the general properties of the photo-assisted amplitude probabilities. \cite{dubois_integer_2013} 

Even more evident is the squeezing behavior of the radiation generated by the Lorentzian drive. Indeed, one has that $\langle B^{2}_{lor} \rangle_{c}$ (black full curve in Fig. \ref{fig2}) goes below both the quantum vacuum and the minimum associated with the cosine drive. This is obtained at the expense of $\langle A^{2}_{lor} \rangle_{c}$ (red full curve in Fig. \ref{fig2}) that is above $\langle B^{2}_{cos} \rangle_{c}$ for the same value of $z$ which minimizes the conjugated quadrature.  Moreover, the mirror symmetry discussed above is absent here. This picture survives also at finite temperature comparable with experiments \cite{gasse_observation_2013} (see Table \ref{tab1}). The present analysis seems to indicate that the strict hierarchy among the periodic drives reported in Refs. \onlinecite{dubois_integer_2013, rech_minimal_2016} for the photo-assisted noise at zero frequency still survives in the finite frequency case (and consequently for the associated electromagnetic quadratures) indicating again the Lorentzian drive as the best candidate in order to minimize the noise (maximize the squeezing). 
\begin{table}
 \begin{tabular}{ | c | c | c | c | }
    \hline
     & $\theta=0$ & $\theta=0.04$ & $\theta=0.08$ \\ \hline
Square & $0.641$ & $0.697$ & $0.753$ \\ \hline 
       Cosine  & $0.618$ & $0.676$ & $0.733$  \\ \hline 
     Lorentzian $(\eta=0.1)$& $0.486$ & $0.549$ & $0.611$ \\ \hline      
    \end{tabular}
    \caption{Minimum of the quadratures for the square, cosine and Lorentzian drive achieved numerically at various experimentally relevant temperatures.}
    \label{tab1}
\end{table}

\begin{figure}[h]
\centering
\includegraphics[scale=0.9]{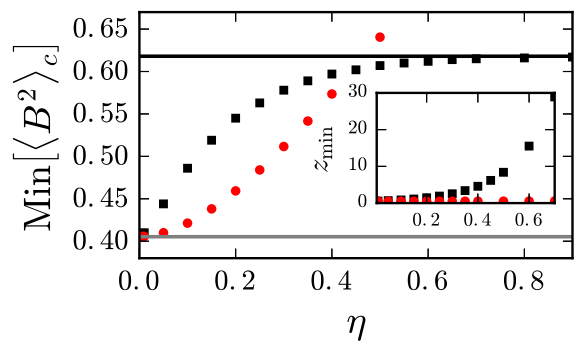}
\caption{(Color on-line) Value of the minimum of $\langle B^{2}_{lor} \rangle_{c}$ (Lorentzian drive, black squares) and $\langle B^{2}_{rect} \rangle_{c}$ (rectangular drive, red circles) in units of $\mathcal{A}$, as a function of $\eta$ and at zero temperature ($\theta=0$). They are compared with the theoretical minimum derived for a Dirac comb $4/\pi^2\approx0.405$ (thin grey horizontal line) and the numerical minimum calculated for the cosine drive ($\approx 0.618$) (thin black horizontal line). The values of $z$ are different for each $\eta$ in order to maximize the squeezing ($z=z_{\mathrm{min}}$). Inset. Corresponding evolution of the value of $z_{\mathrm{min}}$ at which the minimum of the noise occurs (same color code and style for the points).} 
\label{fig3}
\end{figure}

This idea is further strengthened by the analysis of the evolution of the minimum of $\langle B^{2}_{lor} \rangle_{c}$ at zero temperature as a function of the width of the Lorentzian pulse $\eta$ (see black squares in Fig. \ref{fig3}). The minima of the quadrature are obtained by choosing for each $\eta$ the value of parameter $z$ at which the squeezing is maximized ($z_{\mathrm{min}}$). This curve remains above the theoretical minimum $4/\pi^2\approx0.405$ (thin grey horizontal line) predicted in Ref. \onlinecite{mendes_cavity_2015} for a Dirac comb signal and asymptotically approaches the minimum of the cosine drive ($\approx 0.618)$ (thin black horizontal line). \cite{gasse_observation_2013} Notice that, as stated above, the corresponding minimum for the square drive is even higher  ($\approx 0.641$). To complete this analysis, it is also interesting to compare the Lorentzian drive with a train of rectangular pulses (see red circles in Fig. \ref{fig3}). In this case, for a quite extended range of $\eta$ ($0<\eta<0.4$) we have a squeezing which is better with respect to the cosine drive and comparable with the Lorentzian pulse.  Notice that both the Lorentzian and the rectangular pulse converge towards the theoretical minimum discussed above for $\eta \rightarrow 0$, as expected. According to Fig. \ref{fig3} (inset), while the value of $z_{\mathrm{min}}$ increases exponentially with $\eta$ in the Lorentzian case, it stays roughly constant in the rectangular case. However, even if very promising in view of maximizing the squeezing, narrow rectangular pulses requires a greater number of harmonics to be generated experimentally and also present some drawbacks for what it concerns the properties of the emitted quantum photonic states. We will address this point in the following. 

\subsection{Single-photon state purity}
\begin{figure}[h]
\centering
\includegraphics[scale=0.9]{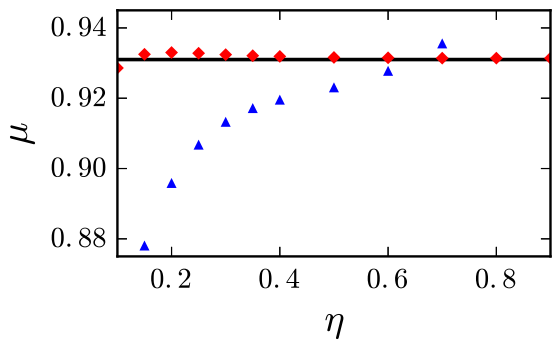}
\caption{(Color on-line) Evolution of the single-photon purity $\mu$ as a function of $\eta$ in the optimal squeezing configuration ($z=z_{\mathrm{min}}$) for a Lorentzian drive (red diamonds) and a rectangular pulse (blue triangles), Curves are compared with the optimal values calculated for a cosine ($\approx 0.931$, thin black horizontal line). The corresponding value for a square wave is obtained for the rectangular pulse at $\eta=1/2$.} 
\label{fig4}
\end{figure}

Another interesting quantity to look at in order to characterize the electromagnetic radiation emitted by mesoscopic devices is the purity of the outgoing single-photon state, defined as \cite{grimsmo_quantum_2016} 
\be
\mu= \frac{1}{2} \frac{|\langle\left[A, B\right] \rangle |}{\Delta A \Delta B}=\frac{\mathcal{A}}{\sqrt{\left(\tilde{\mathcal{S}}-\mathcal{X}^{(1)}\right)\left(\tilde{\mathcal{S}}+\mathcal{X}^{(1)}\right)}}.
\label{Purity}
\ee
This quantity is $\mu=1$ for a pure state and lower than $1$ for a generic mixed state. In Fig. \ref{fig4} we show the behavior of this quantity for the Lorentzian drive (red diamonds) as a function of $\eta$ in comparison with the optimal values obtained for the cosine drive ($\approx0.931$, thin black horizontal line). In all cases we have a value of $\mu$ quite close to one, signature of an highly (even if not perfectly) pure state. Moreover, we can note that there is a region of $\eta$ (very narrow pulses) for which the Lorentzian drive is slightly above the cosine drive. In particular, the purity associated with the Lorentzian drive has a maximum (for $\eta\approx 0.2$) and asymptotically converges to the value of the cosine drive from above. From this result we can deduce the fact that Lorentzian drive can not only generate pure electronic states, namely the Levitons \cite{keeling_coherent_2008, dubois_integer_2013, dubois_minimal_2013, jullien_quantum_2014}, but also quite pure outgoing single-photon squeezed states. For a further comparison, one can see that the purity of the quantum electromagnetic state emitted by a Lorentzian drive is by far greater with respect to the one of a rectangular drive (blue triangles). Indeed, even if we have showed above that this signal leads to a compatible squeezing for the same range of $\eta$, the achieved values of $\mu$ are quite far from the others, approaching 1 only at $\eta>0.7$ where the corresponding squeezing is very small (see Fig. \ref{fig3}). It is worth to note that even higher values of purity, closely approaching unity, can be achieved for the Lorentzian drive case (not shown). However, this regime is reached to the detriment of the squeezing which, even if still present, becomes less effective.

According to these considerations the Lorentzian drive emerges as the most promising candidate among the experimentally feasible drives in order to create pure squeezed photonic states suitable for quantum communication applications. Notice that a recent analysis carried out for an asymmetric quantum dot geometry also indicated a train of Levitons as the candidate to maximize the squeezing of the outgoing electromagnetic field. \cite{mendes_cavity_2015}

\section{Two-photon entanglement}\label{Two_photons}
According to the results derived above and following Ref. \onlinecite{forgues_experimental_2015}, it is possible now to investigate the entanglement properties of microwave photons emitted at different frequencies $\omega_{1}=\omega$ and $\omega_{2}= \omega_{0}-\omega$. Proceeding as before, we can define a quadrature for each frequency, namely 

\beq
A_{1, 2}(\omega_{1,2})&=&\frac{1}{\sqrt{2}}\left[I(\omega_{1,2}) +I(-\omega_{1,2})\right]\\
B_{1, 2}(\omega_{1,2})&=&\frac{i}{\sqrt{2}}\left[I(\omega_{1,2}) -I(-\omega_{1,2})\right].
\eeq

It is then possible to identify various different fluctuations involving these quadratures, which are again connected to different harmonic of the photo-assisted finite frequency noise. 

Due to the constraints $\omega_{1}+\omega_{2}=\omega_{0}$  and $\omega_{1}\neq \omega_{2}$, the diagonal fluctuations read
\beq
&&\langle A^2_{1}  \rangle_{c}=\langle B^2_{1}  \rangle_{c}=\tilde{\mathcal{S}}\\
&&\langle A^2_{2}  \rangle_{c}=\langle B^2_{2}  \rangle_{c}=\tilde{\mathcal{S}}_{\lambda}
\eeq
where $\tilde{\mathcal{S}}_{\lambda}$ represents the photo-assisted noise in Eq. (\ref{S_tilde}) calculated at frequency $\omega_{2}= \omega_{0}-\omega$, but with the parameters properly rescaled with respect to $\omega_{1}=\omega$ in order to allow comparison, namely 

\begin{widetext}
\be
\tilde{\mathcal{S}}_{\lambda}= \frac{1}{2} \sum_{n=-\infty}^{+\infty}P_{n} (z) \left[\mathcal{S}_{0} (q+\lambda-1+n\lambda,\theta)+ \mathcal{S}_{0} (q-\lambda+1+n\lambda,\theta)\right].
\ee
\end{widetext}
The non-diagonal fluctuations can be written instead as
\be
\langle A_{1} A_{2} \rangle_{c}=-\langle B_{1} B_{2}\rangle_{c}= \mathcal{X}^{(1)}.
\ee
It is worth noting that here, differently from what is seen in the previous section, the ratio $\lambda=\omega_{0}/\omega$ is a free parameter \emph{not constrained a priori} by the condition $\lambda=2$.
Concerning the crossed fluctuations one has
\be
\langle A_{1} B_{2} \rangle_{c}= \langle A_{2} B_{1} \rangle_{c}= \mathcal{X}^{(1)}_{+}- \mathcal{X}^{(1)}_{-}
\ee
with, for real probability amplitudes, 
 \beq
 \mathcal{X}^{(1)}_{+}&=&\frac{1}{2} \sum_{n} p_{n}(z) p_{n+1}(z) S_{0}(q+1+n\lambda, \theta)\\
  \mathcal{X}^{(1)}_{-}
  &=&\frac{1}{2}\sum_{m} p_{m+1}(z) p_{m}(z) S_{0}(q-1+\lambda+m\lambda, \theta). 
 \eeq
It is easy to note that, due to the reality of the photo-assisted amplitudes (valid for all the considered drives) and the fact that the sum indices are mute, one has
\be
\langle A_{1} B_{2} \rangle_{c}=  \langle A_{2} B_{1} \rangle_{c}\approx 0
\ee 
as long as $\lambda \approx 2$. 

The above considerations justify the experimental choice $\lambda\approx 2.07$ ($\omega_{1}$ and $\omega_{2}$ different but very close) done in Ref. \onlinecite{forgues_experimental_2015}. Indeed, for this value of the parameter $\lambda$, one can safely neglect the crossed contributions with respect to the others (see Fig. 3 of Ref. \onlinecite{forgues_experimental_2015}).

\subsection{Two-photon squeezing}
It is useful, at this point, to introduce the rescaled operators
\beq
\alpha_{1}&=& \frac{A_{1}}{\sqrt{2 \mathcal{A}}}, \\ 
\alpha_{2}&=& \frac{A_{2}}{\sqrt{2 \mathcal{A} (\lambda-1)}} 
\eeq
and analogous definitions for $\beta_{1}$ and $\beta_{2}$, leading to the quantum limit at zero AC and DC voltage and zero temperature
\be
 \langle \alpha^{2}_{1,2} \rangle_{c}= \langle \beta^{2}_{1,2} \rangle_{c}=\frac{1}{2}.
\ee
It is easy to note that the fluctuations of these new operators are trivially related to the previously discussed photo-assisted noise measured at finite frequency. More interesting are the operators 
\beq
u&=& \frac{\alpha_{1}-\alpha_{2}}{\sqrt{2}},\\
v&=& \frac{\beta_{1}+\beta_{2}}{\sqrt{2}}
\eeq
connecting two photons emitted at different frequencies. Their fluctuations are given by 
\begin{equation}
\langle u^{2} \rangle_{c}=\langle v^{2} \rangle_{c}=\frac{1}{4 \mathcal{A}} \left[\tilde{\mathcal{S}} +\frac{\tilde{\mathcal{S}}_{\lambda}}{\lambda-1}-\frac{2\mathcal{X}^{(1)}}{\sqrt{\lambda-1}} \right].
\end{equation} 
 Proceeding on the same way we can define similar expression also for the combinations $w= (\alpha_{1}+\alpha_{2})/\sqrt{2}$ and $y=(\beta_{1}-\beta_{2})/\sqrt{2}$ (with an opposite sign in front of the non-diagonal contribution). The behavior of the above quantities is qualitatively very similar to what was reported in Fig. \ref{fig2} for the single-photon squeezing (due to the fact that $\lambda\approx 2$ also in this case), but the physical meaning is deeply different. Indeed, they indicate the emergence of non trivial correlations between the fluctuations of the quadratures of photons emitted at different frequencies. This phenomenology usually indicated as \emph{two-photon squeezing} represents a first, even if not conclusive, indication of the possibility of entanglement between the emitted photons. We have that, at low enough temperatures, all the investigated drives show two-photon squeezing at $q\approx 1$ for the operators $\langle u^{2}\rangle_{c}= \langle v^{2} \rangle_{c}$, which is again maximum for the Lorentzian drive case. 

\subsection{Comments about entanglement}
The two-photon squeezing discussed above proves the existence of non-trivial correlations between the quadratures of the electromagnetic field at different frequencies. For strong enough correlations one can have entanglement. This is guaranteed for continuous variables as long as the condition 
\be
\delta= \langle u^{2} \rangle_{c}+ \langle v^{2} \rangle_{c} <1, 
\ee
which constitutes the continuous version of the Bell inequality\cite{duan_inseparability_2000}, is satisfied for some range of parameters. Conversely the two-photon state is separable. The above condition is obviously fulfilled for a relatively wide range of voltages around $q\approx 1$ for all the considered drives and also at experimentally reasonable values of temperature as reported in Table \ref{tab2}. 
\begin{table}
\begin{center}
  \begin{tabular}{ | c | c | c | c | }
    \hline
     & $\theta=0$ & $\theta=0.04$ & $\theta=0.08$ \\ \hline
    Square  & $0.661$  & $0.699$  & $0.752$  \\ \hline 
      Cosine  & $0.639$ & $0.668$ & $0.732$  \\ \hline 
     Lorentzian ($\eta=0.1$) & $0.512$ & $0.551$ & $0.610$  \\ \hline      
         \end{tabular}
\end{center}
\caption{Minimal values of the parameter $\delta$ (at fixed $\lambda=2.07$) achieved numerically for various experimentally relevant drives and temperatures.}
    \label{tab2}
\end{table}

\subsection{Purity of the two-photon state}
Also in this case it is possible to define the purity of the emitted two-photon state. To this aim one can define the quadrature vector  \cite{braunstein_quantum_2016} $\xi= (\alpha_{1}, \beta_{1}, \alpha_{2}, \beta_{2})$
and the related correlation matrix $\gamma_{ij}=2 \langle \xi_{i} \xi_{j} \rangle_{c}$ in such a way that the two-photon state purity reads
\be
\mu_{2}= \frac{1}{\sqrt{Det (\gamma)}}=\frac{(\lambda-1)^{2}\mathcal{A}^{2}}{\tilde{S} \tilde{S}_{\lambda}-\left[\mathcal{X}^{(1)} \right]^{2}}
\ee
extending the definition given in Eq. (\ref{Purity}).

\begin{figure}[h]
\centering
\includegraphics[scale=0.9]{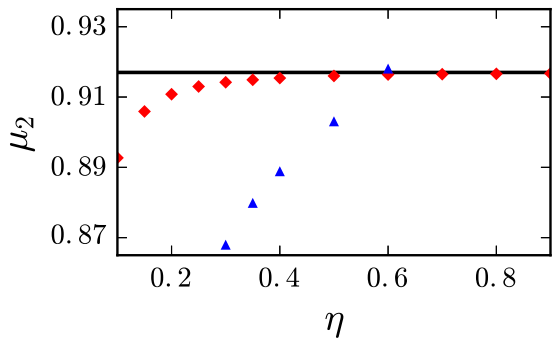}
\caption{(Color on-line) Evolution of the two-photon purity $\mu_{2}$ as a function of $\eta$ in the optimal squeezing configuration ($z=z_{\mathrm{min}}$) for a Lorentzian drive (red diamonds) and a rectangular pulse (blue triangles). They are compared with the optimal values calculated for a cosine voltage ($\approx0.917$, thin black horizontal line).} 
\label{fig5}
\end{figure}

The behavior of this quantity for the various drives is reported in Fig. \ref{fig5}. The behavior is qualitatively similar to what is observed in Fig. \ref{fig4} for the single-photon purity $\mu$, with the only difference that the Lorentzian drive (red diamonds) shows no maximum and asymptotically approaches the value for the cosine drive ($\approx 0.917$, thin black horizontal line) from below. 

The above considerations allows us to drawn interesting conclusions about the prominent role played by the Lorentzian pulses in generating quite pure entangled two-photon states suitable for quantum information application. 

\section{Conclusion}\label{Conclusions}
We have considered a mesoscopic device in a tunnel junction (or QPC geometry) subjected to various different periodic drives. Under this condition the system emits microwave radiation with remarkable non-classical features. We have compared various experimentally relevant drives like cosine, rectangular and Lorentzian in order to determine which one is more suitable in order to enlighten quantum properties of the outgoing electromagnetic field. We have showed that a train of Lorentzian voltage pulses, already investigated at length in the framework of electron quantum optics due to its remarkable peculiarities in terms of single-electron emission, represents the best candidate in order to achieve both single and two-photon squeezing as well as two-photon entanglement in the frequency domain. Moreover, this peculiar drive leads to single and two-photon quantum states of quite high purity. We think that the present analysis will have an important impact on quantum communication applications. Possible further developments of our work could address the role played by interaction in further enhancing or suppressing the observed quantum features of the emitted radiation. \cite{altimiras_dynamical_2014, parlavecchio_fluctuation_2015}

\section*{Acknowledgements}
We are grateful to B. Reulet for useful discussions. The support of Grant No. ANR-2010-BLANC-0412 (``1 shot'') and of ANR-2014-BLANC ``one shot reloaded'' is acknowledged. This work was carried out
in the framework of Labex ARCHIMEDE Grant No. ANR-11-LABX-0033 and of A*MIDEX project Grant No. ANR-
11-IDEX-0001-02, funded by the ``investissements d'avenir'' French Government program managed by the French National
Research Agency (ANR).

\bibliography{biball_2016b}{} 

\end{document}